\documentclass[reprint, 
onecolumn, aps,prb, amssymb, amsmath, superscriptaddress, showpacs, notitlepage] {revtex4-1}
\usepackage{graphicx}
\usepackage{epstopdf}
\usepackage{bm}
\usepackage[colorlinks, linkcolor = blue, citecolor = blue, urlcolor=blue, pdfborder={0 0 0 [0 0]},bookmarks=false]{hyperref}

\begin{document}

\title{H-T Phase Diagram of Rare-Earth -- Transition Metal Alloy
	in the Vicinity of the Compensation Point}

\author{M.~D.~Davydova}
\affiliation{Moscow Institute of Physics and Technology~(State University), 141700 Dolgoprudny, Russian Federation}
\affiliation{Prokhorov General Physics Institute of the Russian Academy of Sciences, 119991 Moscow, Russian Federation}

\author{K.~A.~Zvezdin}
\email{konstantin.zvezdin@gmail.com}
\affiliation{Moscow Institute of Physics and Technology~(State University), 141700 Dolgoprudny, Russian Federation}
\affiliation{Prokhorov General Physics Institute of the Russian Academy of Sciences, 119991 Moscow, Russian Federation}

\author{J. Becker}
\affiliation{Institute for Molecules and Materials, Radboud University, Nijmegen 6525 AJ, The Netherlands}

\author{A.~V.~Kimel}
\affiliation{Institute for Molecules and Materials, Radboud University, Nijmegen 6525 AJ, The Netherlands}
\affiliation{Moscow Technological University (MIREA), 119454 Moscow, Russian Federation}

\author{A.~K.~Zvezdin}
\affiliation{Prokhorov General Physics Institute of the Russian Academy of Sciences, 119991 Moscow, Russian Federation}
\affiliation{Lebedev Physical Institute of the Russian Academy of Sciences, 119333 Moscow, Russian Federation}
\email{zvezdin.ak@phystech.edu }

		\begin{abstract}
			Anomalous hysteresis loops of ferrimagnetic amorphous alloys in high magnetic field and in the vicinity of the compensation temperature have so far been explained by sample inhomogeneities. We obtain H-T magnetic phase diagram for ferrimagnetic GdFeCo alloy using a two-sublattice model in the paramagnetic rare-earth ion approximation and taking into account rare-earth (Gd) magnetic anisotropy. It is shown that if the magnetic anisotropy of the $f$-sublattice is larger than that of the $d$-sublattice, the tricritical point can be at higher temperature than the compensation point. The obtained phase diagram explains the observed anomalous hysteresis loops as a result of high-field magnetic phase transition, the order of which changes with temperature. It also implies that in the vicinity of the magnetic compensation point the shape of magnetic hysteresis loop is strongly temperature dependent.	
		\end{abstract} 	

	\maketitle

	\section{Introduction}

		Rare-earth amorphous alloys and intermetallics is a large class of magnetic materials allowing to change their magnetic properties in a wide range by a subtle change of the composition, temperature or application of magnetic field \cite{buschow1977intermetallic,franse1993magnetic,belov1976spin,duc2002metamagnetism,sechovsky1994giant}. The materials have already found applications as hard magnets or
		recording media and they still offer a rich playground in the areas of spintronics \cite{RevModPhys.76.323}, magnonics \cite{kruglyak2010magnonics} and ultrafast magnetism \cite{PhysRevLett.99.047601,PhysRevLett.103.117201,moser2002magnetic, ostler2012ultrafast,PhysRevLett.108.127205,radu2011transient,graves2013nanoscale,PhysRevLett.98.207401,le2012demonstration,doi:10.1063/1.3339878}.

		GdFeCo is a particular example of such amorphous alloys. It is a 3$d$-4$f$ ferrimagnet with compensation temperature \cite{PhysRevB.22.1320}, at which the magnetizations of the two sublattices become equal. At temperatures lower than the compensation temperature, the magnetization of the rare-earth (Gd) sublattice $M_f$ is larger than that of the transition metal (Fe) $M_d$ ($M_f - M_d >0$), whereas at higher temperatures $M_f - M_d <0$. Many studies of GdFeCo, GdFe and GdCo compounds as well as magnets with different rare-earth ion in high magnetic field revealed triple hysteresis loops in the vicinity of the magnetization compensation point \cite{Esho_1976, CHEN1983269, OKAMOTO1989259, amatsu1977anomalous,PhysRevLett.118.117203}. The observed triple loops are clearly different from a hysteresis loop normally expected for a single thin film, where one would not expect a sudden decrease in magnetization in the strong applied magnetic field.  However, hysteresis loops of this form are typical for multilayered structures. To emphasize the difference, we will refer to the the loops in single-layer structures as anomalous. These loops are strongly dependent on temperature. Earlier similar behavior was explained by sample inhomogeneities \cite{amatsu1977anomalous} or strong exchange bias between surface and bulk layers that have different stochiometric composition of the alloy; in particular, this lead to estimation of the strongest ever reported exchange bias field of several Tesla \cite{chen2015observation}. However, to date no theoretical model has been proposed that would allow to calculate magnetization curves that would explain experimental data. Here we use a model for a homogeneous two-sublattice ferrimagnet film and suggest an alternative explanation for the observed anomalous hysteresis loops.

\begin{figure}[htb]%
			\includegraphics[width=0.35\columnwidth]{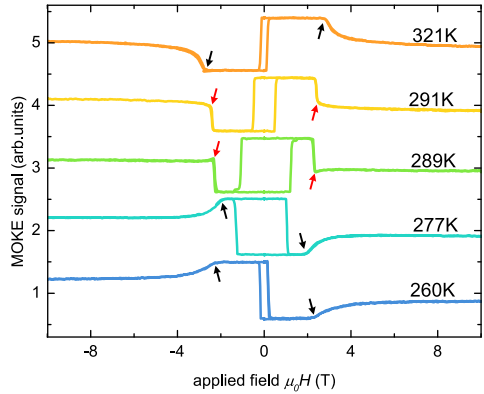}%
			\caption{%
				Static
				magneto-optic Kerr effect in GdFeCo sample measured at 630 nm probe
				wavelength at different temperatures from 260 to 321 K. A
				paramagnetic background was subtracted from the
				measurements. The compensation temperature is 283 K. Ref. \cite{PhysRevLett.118.117203}. Black and red arrows indicate second- and first-order transitions, respectively. }
			\label{fig1}
		\end{figure}

		Figure \ref{fig1} shows the results of high magnetic field measurements of the magneto-optical Kerr effect in GdFeCo \cite{PhysRevLett.118.117203}. The composition of the alloy with 24 $\%$ Gd, 66.5 $\%$ Fe and 9.5 $\%$ Co resulted in the compensation temperature $T_{comp}$=283 K. The field was applied at the normal to the sample, which is also the easy magnetization axis. The measurements were done at the probe wavelength of 630 nm in the polar Kerr geometry. In this case the probe is predominantly sensitive to the magnetization of the Fe-sublattice. Therefore the obtained hysteresis loops reveal the field dependence of the orientation of the Fe-magnetization.
		It is seen upon an increase of the field first a minor hysteresis loop shows up, which corresponds to the magnetization reversal. Further increase of the field does not change the orientation of the magnetization until a critical field is reached. This field launches spin-flop transition which is seen as a decrease of the magneto-optical signal. At this field the magnetizations of the sublattices turn from the normal
		of the sample, get canted and form a non-collinear state.
		The character of the spin-flop transition changes with 
		temperature. 
		Below the magnetic compensation temperature the spin-flop transition occurs gradually (see loops for 260 K and 277 K in Fig. 1). Just above the compensation point at the spin-flop field one observes abrupt change in the magnetic structure (see loops for  289 K and 291 K in Fig. 1). Upon further increase of the sample temperature the transition is seen as gradual again (see loop at 321 K in Fig. 1). 
		Abrupt and gradual changes in magnetization induced by external magnetic field are characteristic features of first- and second-order phase transitions, respectively. Hence, these 
		measurements imply that the order of the phase transition changes from second to first and back to second upon a temperature increase across the compensation point. Such a temperature-dependent order of the spin-flop transition has not been described for GdFeCo in literature before.
		
		 Note that, although phase diagrams for 3$d$-4$f$ ferrimagnets were first obtained theoretically almost 50 years ago \cite{goransky1970temperature,zvezdin1972some} and supported by numerous experiments (see [\onlinecite{zvezdin1995field}] and references therein), in the studies performed so far the anisotropy of the transition metal sublattice was taken to be larger than that of the rare-earth sublattice. The existing results on the magnetic phase diagrams fail to explain the anomalous hysteresis loops observed experimentally \cite{amatsu1977anomalous,PhysRevLett.118.117203,chen2015observation}. Unusual behavior of the critical fields in rare-earth intermetallics in the case of prevailing anisotropy of the rare-earth sublattice was recently investigated by some of co-authors theoretically for HoFe$_x$Al$_{12-x}$ \cite{sabdenov2017magnetic,sabdenov2017magnetic1}.
		 Here we show that if the rare-earth anisotropy is larger than that of the transition metal, the tricritical point on the phase diagram lies at higher temperatures with respect to the compensation point. As a result, the observed hysteresis loops can be explained in terms of intrinsic first- and second-order phase transitions in the intermetallic samples.

		\section{Magnetic phase diagram}

		To obtain the $H$-$T$ phase diagram, we derive the thermodynamic potential for a two-sublattice ferrimagnet in paramagnetic rare-earth ion approximation \cite{goransky1970temperature,zvezdin1995field,zvezdin1972some}. We start with the Hamiltonian for a system of $f$- and $d$- ions in external magnetic field in the form \cite{zvezdin1985rare}:
		\begin{equation} \label{H}
		\begin{split}
		\mathcal H = \mathcal H_f + \mathcal H_{f-d} + \mathcal H_d, 
		\end{split}	
		\end{equation}

		\noindent where 
		
		\begin{equation} \label{Hdetails}
		\begin{split}
		&\mathcal H_{d} = \mathcal H^{d}_{cr} -  \sum _{i_1, i_2 \in d} \mathcal{J}^{d}_{i_1 i_2} \bm S _{i_1}  \bm S_{i_2} + |g_J^{d}| \mu_B \bm H \sum_{i \in d } \bm J_{i},\\
		&\mathcal H_{f-d} = - \sum_{i_1\in f, i_2 \in d }\mathcal{J}^{f-d}_{i_1 i_2} \bm S _{i_1}  \bm S_{i_2},
		\\
		&\mathcal H_{f} = \mathcal H^{f}_{cr}  + |g_J^{f}| \mu_B \bm H \sum_{i \in f } \bm J_{i},
		\end{split}	
		\end{equation}

		\noindent In one-sublattice Hamiltonian $\mathcal H_{d}$ for $d$-sublattice the first term represents the crystal field Hamiltonian (see Appendix A), the second term is intra-sublattice exchange interactoin and the last term is the Zeeman energy in the external magnetic field $\bm H$. The second component $H_{f-d}$ of the total Hamiltonian  is the intrasublattice exchange interaction. The $f$-sublattice Hamiltonian $\mathcal H_{d}$ consists of crystal field and Zeeman energy. We neglect the exchange within $f$-sublattice because its magnitude is several orders smaller than $f-d$ exchange \cite{zvezdin1985rare}. The summation is performed over the ions belonging to $f$ and $d$ sublattices, $\bm J_{i} = \bm L_i + \bm S_i$ is the total angular momentum of an operator for the $i$-th ion, $\mathcal J^{d}$ and $\mathcal J^{f-d}$ are the matrices of the exchange interaction within one sublattice and between sublattices, correspondingly. In the following, we assume the g-factors for rare-earth and transition metal $s$-ions to be $g \equiv |g^f_J| \approx|g^d_J|  \approx 2 $. 
		
		Using the procedure described in Appendix A we derive the thermodynamic potential of nonequilibrium state (effective free energy) where the parameter is the orientation of the $d$-sublattice magnetization vector $\bm M_d$. The value of this magnetization is assumed to be saturated due to the large $d$-$d$ exchange with corresponding exchange field of the order of $10^6-10^7$ Oe. We also assume that the magnetization of the $f$-sublattice is defined by the effective magnetic field acting on it $\bm H_{eff} = \bm H - \lambda \bm M_d$, where $\lambda$ is the $f$-$d$ exchange coupling constant (see Appendix A).  Finally, we arrive at the thermodynamic potential in the form given by eq. \eqref{th}. Finally, we obtain:
		\begin{equation} \label{th}
			\Phi = -\bm M_d \cdot \bm H - \int_0^{H_{eff}} g J B_J(\frac{g J \mu_B h }{k T}) \mathrm{d} h + K_f \sin^2 \theta_f + K_{d} \sin^2 \theta_{d},
		\end{equation}
		where $B_J(x)$ is the Brillouin function, $J = 7/2$ is the ground state total angular momentum of Gd ion, and $K_f$, $K_d$ denote the uniaxial magnetic anisotropy constants for the two sublattices, which are assumed to have different values. In our spherical coordinate system, the polar axis lies in the direction of the easy magnetization axis, and the angles $\theta_f$ and $\theta_d$ are the polar angles for magnetizations of rare-earth and transition metal sublattices, respectively.

		When the magnetic field $\bm H$ is applied along the easy axis, the effective free energy may be represented as a function of the single order parameter $\theta_d$:
		\begin{equation} \label{conv}
			\begin{split}
			\Phi = - M_d H \cos \theta_d - \int_0^{H_{eff}(\theta_d)} g J B_J\left( \frac{g J \mu_B h }{k T}\right) \mathrm d h + K_f \left(\frac{\lambda M_d \sin \theta_d}{H_{eff}(\theta_d)} \right)^2 + K_{d} \sin^2 \theta_{d},
			\end{split}	
		\end{equation}
		where $H_{eff}(\theta_d)=\sqrt{H^2 + \lambda^2 M_d^2 - 2 H \lambda M_d \cos \theta_d}$.

	Using the expression for the thermodynamic potential \eqref{conv} and the method described in Ref. \cite{zvezdin1995field} we numerically calculate the magnetic phase diagram in the coordinates '$H$-$T$' (Fig. \ref{fig}). The ground states of the system are found by minimization of the thermodynamic potential \eqref{conv} with regard to the order parameter $\theta_d$. At the minima one finds $\frac{\partial \Phi}{\partial \theta_d} =0$ and $\frac{\partial^2 \Phi}{\partial \theta _d^2} >0$. 	The lines of stability loss, where $\frac{\partial^2 \Phi}{\partial \theta _d^2} = 0$, are found for each phase. In terms of Landau theory of the phase transitions, if the thermodynamics potential is written in terms of Taylor series with respect to the order parameter $\Phi= a(H,T) \theta_d^2 + \frac{1}{2} b(H,T) \theta_d^4 + c(H,T) \theta_d^6+...$, the second-order phase transition is observed when $ a(H,T) =0$ and $b(H,T)$ is positive. If $a(H,T)>0$, $c(H,T)>0$, but $b(H,T)<0$, the system undergoes the first-order phase transition. 
 	Near the first-order phase transition two possible stationary states coexist, corresponding to one local (metastable) and one global (stable) minimum of the thermodynamic potential, respectively. 
 	
	For the numerical calculations, we used the following set of parameters: $T_{comp}$ = 283 K, $T_{C}$ = 500 K, $M_f(0) = 7 $ $\mu_B$/f.u., $M_d(0) = 6.5 $ $\mu_B$/f.u., where f.u. means 1 formula unit, and the exchange constant $\lambda = 22$ T/$\mu_B$. To the best of our knowledge, no experimental data about the strength of the magnetic anisotropy of the rare-earth sublattice is available for GdFeCo alloy. Nevertheless, until now it has been believed that the magnetic anisotropy of the
	 Gd-sublattice is smaller than the one of the iron sublattice. Here we show that if $K_f - K_d \approx 0.6$ K/f.u. one obtains a qualitative agreement of the calculated magnetic phase diagram with the experimental data from the recent study \cite{PhysRevLett.118.117203}.

	For analytical investigation of the phase diagram, we describe the two- sublattice ferrimagnet in terms of the antiferromagnetic vector $\bm L = \bm M_f- \bm M_d$ and the net magnetization $\bm M = \bm M_f + \bm M_d$. Note that in the vicinity of the compensation point the difference between the sublattice magnetizations $|M_f-M_d |\ll L$ is small but not zero. These two vectors are parametrized using sets of angles $\theta, \varepsilon$, $\varphi$ and $\beta$. The angles are defined so that:
	\begin{equation}
	\begin{split}
	\theta_f &= \theta -\varepsilon, \ \theta_d = \pi - \theta - \varepsilon, \\
	\varphi_f &= \varphi - \beta, \ \varphi_d = \pi + \varphi + \beta,
	\end{split}	
	\end{equation}
	where $\varphi_f$ and $\varphi_d$ are the azimuthal angles for magnetizations of rare-earth and transition metal sublattices, respectively. 
	In the chosen coordinate system the azimuthal axis lies in plane perpendicular to the easy axis. In this case the antiferromagnetic vector may be naturally defined as $\bm L = (L \sin \theta \cos \varphi, L \sin \theta \sin \varphi, L \cos \theta)$. In the vicinity of the second-order phase transition the expansion of the thermodynamic potential \eqref{th} may be performed in the series of angles $\theta$, $\varepsilon$, $\varphi$, and $\beta$, which can be seen as the order parameters. Using the expansion, we obtain analytical expressions that describe the behavior of the order parameters in different phases in the vicinity of the compensation temperature. 
	
	In the collinear phase to the left from the compensation point (green area in Fig. \ref{fig}(a)) the parameter $\theta$ is equal to $0$. To the right from the compensation temperature a collinear phase with $\theta = \pi$ is the stable phase (blue area in Fig. \ref{fig}(a)). The noncollinear phase, which is shown in Fig.~\ref{fig}(a) by yellow area can be described by analytical expression: 
	\begin{equation} \label{th1}
	\begin{split}
	\cos \theta \approx -\frac{\chi \left( H^2 - H_1^2 \right) }{6 \chi H \Delta H_{A}} \pm
	\frac{ \sqrt{\chi^2 \left( H^2 - H_1^2 \right)^2 + 12 \chi H^2 \Delta H_{A} \left[ \left(M_f-M_d \right) + \chi \Delta H_A \right] } }{6 \chi H \Delta H_{A}},
	\end{split}
	\end{equation}
	where $\chi = \left( M_f + M_d \right) ^2/(2 M_d M_f \lambda)$, $H_1^2 = 2\frac{K_d + K_f}{\chi}$ and $\Delta H_A = 2\frac{K_f - K_d}{M_f + M_d}$. If the condition $\Delta H_A>0$ is satisfied, the first-order transition between the non-collinear and the collinear phase $\theta = \pi$ will occur at temperatures higher than the compensation point, which follows from expression \eqref{th1}.

	\section{Results and discussion}

	\begin{figure}[htb]%
		\includegraphics[width=0.5\columnwidth]{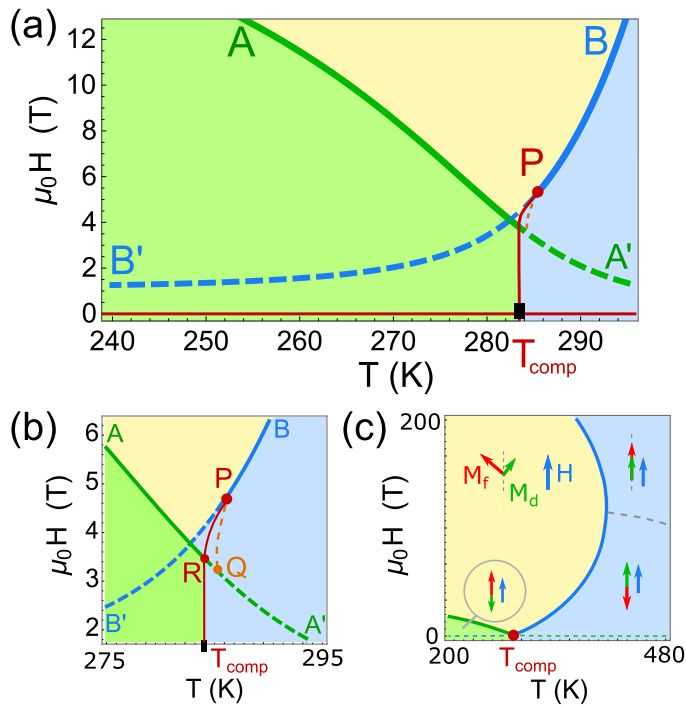}%
		\caption{%
			(a) The $H$-$T$ phase diagram near the compensation point $T_{comp}$ for GdFeCo in magnetic field directed along the easy magnetization axis. 
			The dashed lines are the stability loss lines for the corresponding phases, solid lines correspond to the second-order phase transition, and the line $PQ$ corresponds to the stability loss of the non-collinear phase. 
			(b) The magnified area of the $H$-$T$ phase diagram near the tricritical point $P$. 
			(c) The qualitative zoomed out $H$-$T$ phase diagram up to magnetic fields of the order of 200 T. 
		}
		\label{fig}
	\end{figure}

	Three different phases are present in the magnetic phase diagram. Figure \ref{fig} shows these phases: low-temperature collinear $\theta = 0$ (green area), high-temperature collinear $\theta= \pi$ (blue area) and the non-collinear (angular) phase $\theta=\theta(H,T)$ (yellow area), which is described by Eq. \eqref{th1}. The collinear phase $\theta = 0$ exists below $AA'$ line, whereas the collinear phase $\theta = \pi$ exists below $BB'$ line. These lines are the stability loss lines for the corresponding phases. The area of the angular phase is limited from below by $AQPB$-curve. The zoomed in area of the phase diagram around the point $P$ is shown in Fig. \ref{fig}(b) and the zoomed out phase diagram is shown in Fig. \ref{fig}(c) along with schematically drawn directions of the sublattice magnetizations in each phase. At the dashed gray line in Fig. \ref{fig}(c) the condition $H_{eff} (T) = 0$ is fulfilled. 

	There are several first- and second-order phase transitions  in  the vicinity of the magnetization compensation temperature $T_{comp}$. The second-order phase transitions are denoted by lines $AR$ and $PB$ and characterized by a continuous change of the order parameter across the line. 
	The dashed lines in Fig. \ref{fig} are the lines of the stability loss and denote the theoretical temperature-dependent boundaries for the field hysteresis around the first-order phase transition at the line $H=0$. 
	 The line between $T_{comp}$ and point $P$ corresponds to another and less trivial first-order phase transition. 
	Magnified area of the magnetic phase diagram in the  vicinity of the point $P$ is shown in Fig. \ref{fig}(b). The line $T_{comp}R$ corresponds to the line at which the two collinear phases phases ($\theta = 0$ and $\theta = \pi$) have equal thermodynamic potentials $\Phi(0) = \Phi(\pi)$. Both phases coexist to the left and to the right from $T_{comp}R$. Above line $AR$ there is no minimum of the thermodynamic potential for the collinear phase $\theta = 0$ anymore and the spins turn continuously into the non-collinear phase. At line $RP$ the first-order phase transition continues, but now it is the transition between the angular phase and the collinear $\theta = \pi$ phase. At point $P$ the order of the transition changes from first to second. According to the conventional classification, this is the tricritical point \cite{lawrie1984phase}, in the vicinity of which many physical quantities, such as heat capacity or magnetic susceptibility, experience anomalous behavior.

		\begin{figure}[htb]%
		\includegraphics[width=0.35\columnwidth]{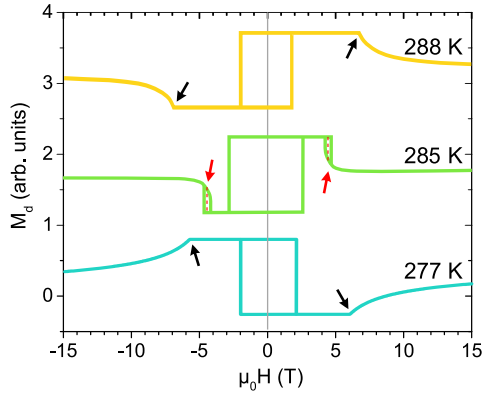}%
		\caption{%
			Dependence of the component of the $d$-sublattice magnetization along the easy axis direction on the magnetic field at different temperatures. Black and red arrows indicate second- and first-order transition points, respectively. }
		\label{fig2}
	\end{figure}

	The first-order phase transition across $T_{comp}P$ line and the tricritical point $P$ in rare-earth ferrimagnets with similar properties were reported earlier	\cite{goransky1970temperature,zvezdin1972some}. However, in the previous studies it was claimed that the temperature corresponding to the tricritical point is smaller than the magnetization compensation temperature $T_P<T_{comp}$.	
	The possibility for anomalous temperature dependent hysteresis loops in  the vicinity of the compensation point of ferrimagnets had been overseen and it had been believed that the observed hysteresis loops are due to inhomogeneities. 
	The relation between this previously overseen first-order phase transition and the observed hysteresis behavior is  as follows. Applying an external magnetic field and measuring the magnetization behavior, one expects to observe a minor hysteresis loop corresponding to the first-order phase transition between two collinear phases $\theta=0$ and $\theta= \pi$. The coercive field of this minor hysteresis loop increases upon approaching the compensation temperature. In the temperature range between the compensation and the tricritical points ($T_{comp}<T<T_P$) upon an increase of the external magnetic field the compound undergoes not one, but two first-order phase transitions. First - the one, which results in a hysteresis loop around $H=0$, as explained above. Second - the spin-flop transition to the non-collinear phase, which will also result in a hysteresis at higher magnetic fields. 
	The size of the second jump of magnetization and its hysteresis will then decrease and, subsequently, vanish at the tricritical point.
	Figure \ref{fig2} shows the calculated magnetic field dependencies of the normal component of the $d$-sublattice magnetization at various temperatures in the vicinity of the compensation point. One can see a remarkable qualitative agreement of the calculations with anomalous temperature dependent hysteresis loops earlier observed in rare-earth transition metal alloys experimentally. 
	Hence, here we have suggested an alternative explanation of the anomalous hysteresis loops without relying on inhomogeneities and large exchange-bias field. The observed hysteresis loops can be seen as an intrinsic property and explained in terms of first- and second-order phase transitions in the compound.

	In the last decade the spin dynamics of rare-earth transition-metal alloys has been attracting an intense research interest due to the unique capability of these materials to reverse their magnetization at the record-breaking speed under action of sub-picosecond laser pulses \cite{PhysRevLett.99.047601}.
	In the research aiming to understand the mechanisms of the ultrafast laser-induced magnetization reversal computational methods have been playing a decisive role \cite{PhysRevLett.103.117201, ostler2012ultrafast,radu2011transient,PhysRevB.87.224417,PhysRevB.96.014409,PhysRevB.92.094411}. It is clear that the value of the magnetic anisotropy of the rare-earth sublattice in ferrimagnets is an important input parameter which may greatly influence the outcome of such simulations. 
	In GdFeCo, the rare-earth anisotropy constant may be expected to be larger than that of iron because the strength of the spin-orbit coupling depends on the nucleus charge $Z$ very close to $Z^4$-law (for more accurate evaluations, see Refs. \cite{blume1963theory,PhysRev.134.A320}). Taking into account excited multiplets with nonzero orbital angular  momentum $L$,  the large single-ion anisotropy can be explained as a result of the spin-orbit coupling and the crystal field. 
	More specifically, the large rare-earth anisotropic contribution can be calculated from microscopic theory by taking into account local crystal field of single rare-earth ion environment and spin-orbit coupling simultaneously: $\hat V = \sum_i \lambda_{SO} \hat {\bm l_i} \hat {\bm s_i} + \sum_{i} \sum_{k,q} B^q_k \hat C^k_q( {\bm l_i}) $, where $\lambda_{SO}$ is the spin-orbit coupling constant, index $i$ spans $f$-electrons of Gd$^{3+}$ ion, $B^q_k$ are the crystal field parameters and $\hat C^k_q( {\bm l_i})$ are the irreducible tensor operators. In perturbation theory of the third order and by taking into account states from both ground $^8S$ and excited $^6P$ terms one obtains the spin-hamiltonian with contribution of the form $D \left( S^2_z - \frac{1}{3} S (S+1)\right)$ \cite{van1934lxxxiii,watanabe1957ground,hutchison1957paramagnetic}. The existing estimations of $D$ from both theory and experiment \cite{al2013electron} are of the order of $10^{-2}-10^{-1}$ cm$^{-1}$/ion. Such a value corresponds to the large gadolinium anisotropy constant $K_R$ used in our calculations.

	 Moreover, it is expected that in compounds with rare-earth ions with non-zero orbital momentum in the ground state (Tb, Dy, Sm), the effect of the rare-earth magnetic anisotropy will be even more pronounced than in the case of Gd. 
	For instance, in the simulations of TbCo \cite{PhysRevB.96.014409} in order to mimic the experimentally observed dependence of magnetic anisotropy on concentration of Tb, it was necessary to set 10 times larger anisotropy for the Tb subllatice compared to the one of Co. Our work provides an approach for experimental verification of element-specific magnetic anisotropies in the rare-earth transition metal ferrimagnets.

	\section{Conclusion}
	
	In conclusion, we investigated the $H$-$T$ phase diagram for a rare-earth - transition metal ferrimagnet in the case of magnetic field directed along the easy magnetization axis. 
	We showed that 
	if the rare-earth anisotropy is larger than that of the $d$-sublattice, the spin-flop transition from collinear to noncollinear phase is either the first- or the second-order phase transition. 
	Just above the compensation temperature the phase transition is of the first order. Starting from the tricritical  point $P$, and at higher temperatures the spin-flop becomes a phase transition of the second order. 
	Such a temperature dependent order of the transition from collinear to non-collinear spin phase allows us to explain anomalous hysteresis loops in rare-earth-transition metal alloys without involving the exchange bias between the surface and the bulk. Hence, we suggest that such hysteresis loops are an intrinsic property of the alloys of GdFeCo-type, which have become model materials in spintronics \cite{dai2012spin}, magnonics \cite{PhysRevLett.98.207401,PhysRevB.73.220402} and ultrafast magnetism 
	\cite{PhysRevB.65.012413,doi:10.1063/1.3119313,PhysRevLett.99.217204,graves2013nanoscale}.
	Note that at the tricritical point many response functions (heat capacity, magnetic susceptibility, etc.) experience anomalous behavior, which open totally new opportunities for fundamental and applied research of the alloys.

	\section{Acknowledgements}

This research has been supported by RSF grant No. 17-12-01333.


	\appendix{
		\section{Derivation of the thermodynamic potential}
	
	We start from a more general form of the Hamiltonian introduced in eq. \eqref{H} that includes exchange interaction within $f$-sublattice. This term can often be neglected due to its smallness\cite{zvezdin1985rare}. First, we restrict ourselves to a ground state term and use Wigner-Eckart theorem to express the spin operators $\bm S_{i}$ through total mechanical momentum $\bm J_{i}$. We obtain the components of the total Hamiltonian:
	
		\begin{equation} \label{H1}
	\begin{split}
	 \mathcal H_{f,d} &= \mathcal H^{f,d}_{cr} - \frac{1}{2} (g_J^{f,d})^2 \mu_B^2 \sum _{i_1, i_2 \in f,d} \bm J _{i_1} T^{f,d}_{i_1 i_2} \bm J_{i_2} + |g_J^{f,d}| \mu_B \bm H \sum_{i \in f,d } \bm J_{i},\\
	\mathcal H_{f-d} &= - g_J^d g_J^f  \mu_B^2 \sum_{i_1\in f, i_2 \in d }\bm J _{i_1} T^{f-d}_{i_1 i_2} \bm J_{i_2},
	\end{split}	
	\end{equation}
	
	\noindent where exchange matrices $T_{d,f}$ and $T_{f-d}$ are linearly proportional to those of $\mathcal J^{d,f}$ and $\mathcal J^{f-d}$. We introduce an effective free energy (the thermodynamic potential of nonequilibrium state, see ref. [\onlinecite{zvezdin1985rare}]) that is the function of both magnetic field and magnetizations $\bm M_{f,d}$:

	\begin{equation} \label{F}
	\Phi = F + \bm h_f \bm M_f + \bm h_d \bm M_d,
	\end{equation}
	
	\noindent where 
	
	\begin{equation}
	F = - T \ln \mathrm{Tr}\left\lbrace  \exp \left( \mathcal{H} - \bm h_f \hat{\bm M}_f - \bm h_d \hat{\bm M}_d\right)  \right\rbrace
	\end{equation}

 	\noindent is the thermodynamic free energy and the total sublattice magnetization operators are  $\hat{\bm M}_{f,d} = g \mu_B  \sum_{i \in f,d} \bm J_i$. The equations for the sublattice magnetizations $\bm M_{f,d} = - \frac{\partial F}{\partial \bm h_{f,d}} $ are viewed as the conditions defining the values of Lagrange multipliers $\bm h_f$ and $\bm h_d$.  In the derivation, we take the trace over the ground state terms, whereas tracing for the excited states may account for a large rare-earth ion anisotropy. This question was discussed above in the paper. 

	Using that the intersublattice exchange energy is 2-3 orders of magnitude smaller than the exchange within the $d$ subsystem we assume the $f$-$d$ homogeneous Hisenberg exchange being equal to 

	\begin{equation}
	\mathcal H_{f-d} = \widetilde{\mathcal {H}}_{f-d}=- \frac{1}{N_d 	} g  \mu_B \sum_{i_1 \in f, i_2 \in d  }\bm J _{i_1} T^{f-d}_{i_1 i_2} \bm M_d  \end{equation}
	
	 \noindent and treat the $d$-subsystem in the mean-field approximation \cite{zvezdin1985rare}. From this equation, the $f-d$ exchange coupling constant $\lambda$ can be determined. 
	 
	 In our approximation, the absolute value of the magnetization $\bm M_d$ is saturated by the $d$-exchange and only its direction is varying. The matrix elements of the  crystal field Hamiltonian are small in comparison to both exchanges, thus we can treat it perturbatively; we also neglect the $f$-$f$ exchange. We obtain:
	
	\begin{equation} \label{F2}
	\Phi = - \bm M_d \cdot \left( \bm H - \bm h_d\right) +\left\langle \mathcal H_{cr}^d \right\rangle  - T \ln \mathrm{Tr}_f\left\lbrace  \exp \left( \mathcal{H}_f + \widetilde{\mathcal {H}}_{f-d} - \bm h_f \hat{\bm M}_f \right)  \right\rbrace + \bm h_f \bm M_f,
	\end{equation}
	
	\noindent where $\mathrm{Tr}_f$ denotes the trace over the $f$-subsystem ground state term states. This is a quite general result that allows for a high accuracy treatment of $f$-$d$ magnets. For subsequent consideration we simplify this expression further. For a GdFeCo-like alloy the single-ion crystal field for both sublattices may be represented by its first term of expansion $\left( \mathcal H^{f,d}_{cr}\right) _i =   \left( A^2_0\right) _i \sum_j C^0_2(\bm l_j)   $, where $C^n_m$ are the Steven's operators \cite{abragam2012electron} and $\bm l_j$ is the angular momentum of $j$-th electron belonging to $i$-th ion. According to Wigher-Eckart theorem if we restrain our consideration to the ground state term with given $J$, the result can be represented as a function of total angular momentum of the ion: $\left( \mathcal H^{f,d}_{cr}\right) _i  = \left( B^2_0\right) _i	 Y^0_2(\bm J_i) $. When viewing the crystal field as a perturbation, we introduce the quantization axis along the external field and find $d$-sublattice 
	$ \left\langle \mathcal H^{d}_{cr} \right\rangle   = - 4/3 \sqrt{\pi/5}  K_d Y^0_2(\bm M_d/M_d )   $ \cite{abragam2012electron}, where $Y^0_2$ are the spherical harmonics and we have introduced the uniaxial magnetocrystalline anisotropy  $K_d$.

	Treating the crystal field acting on $f$ ions as perturbation (similarly to $d$-crystal field), we also assume the magnetization $\bm M_f$ to be aligned with the effective magnetic field acting on it and release the Lagrangian multiplier $\bm h_f$, obtaining the Brillouin function  for after tracing the third term in expression \eqref{F2}: $ M_f \approx g J_f \mu_B N_f B_J \left( \frac{g J_f \mu_B  {H}_{eff}}{k T}\right) $, where $\bm H_{eff} = \bm H - \lambda \bm M_d$ and the $f$-$d$ exchange coupling constant $\lambda = \mathcal J$. The total angular momentum eigenvalue $J_f$ for the ground state term $^8 S$ of Gd ions is equal to 7/2. Finally, we arrive at the thermodynamic potential in the form given by eq. \eqref{th}. }
	
	%

%
\end{document}